\documentclass[aps,pre,twocolumn,groupedaddress,showpacs]{revtex4-1}

\usepackage{natbib} 
\usepackage{amsfonts}
\usepackage{amsmath}
\usepackage{amssymb}
\usepackage[american]{babel}
\usepackage[T1]{fontenc}
\usepackage{graphicx}
\usepackage{color}

\newcommand{\eps}{\mathcal{E}}

\newcommand{\gwlc}{{\sc Gwlc}}

\begin{document}


\title{A minimal model for the inelastic mechanics of biopolymer
  networks and cells}


\author{Lars Wolff}
\author{Klaus Kroy}
\email[]{klaus.kroy@uni-leipzig.de}
\affiliation{Institut f\"ur Theoretische Physik, Universit\"at Leipzig, Postfach 100920, 04009 Leipzig, Germany }


\date{\today}

\begin{abstract}
  Live cells have ambiguous mechanical properties. They were often
  described as either elastic solids or viscoelastic fluids and have
  recently been classified as soft glassy materials characterized by
  weak power-law rheology.  Nonlinear rheological measurements have
  moreover revealed a pronounced inelastic response indicative of a
  competition between stiffening and softening.  It is an intriguing
  question whether these observations can be explained from the
  material properties of much simpler {\it in-vitro} reconstituted
  networks of biopolymers that serve as reduced model systems for the
  cytoskeleton. Here, we explore the mechanism behind the inelastic
  response of cells and biopolymer networks, theoretically. Our
  analysis is based on the model of the inelastic glassy wormlike
  chain that accounts for the nonlinear polymer dynamics and transient
  crosslinking in biopolymer networks. It explains how inelastic and
  kinematic-hardening type behavior naturally emerges from the
  antagonistic mechanisms of viscoelastic stress-stiffening due to the
  polymers and inelastic fluidization due to bond breaking. It also
  suggests a simple set of schematic constitutive equations which
  faithfully reproduce the rich inelastic phenomenology of biopolymer
  networks and cells.
\end{abstract}

\pacs{87.16.ad, 83.60.-a, 83.10.-y}

\maketitle



\section{Introduction} 
The mechanical properties of cells strongly influence processes in
living organisms on all length scales, from the locomotion of single
cells to the expansion of whole tissues during breathing. Therefore,
studying the material properties of cells has a long tradition
\cite{Pelling2008}. One of the major lessons learned so far is that
cells are neither solid nor fluid, but can tune their mechanical state
according to their needs. It is clear by now that this mechanical
state is not simply predefined by biochemical processes, but that
cells can be characterized as complex materials with nontrivial
nonlinear mechanical feedback. Cell mechanics also couples back to the
physiological response of cells \cite{Fletcher2010}, for example to
the spreading behavior \cite{Pelham1997} or even to stem cell
differentiation \cite{Discher2005}. In this contribution, we aim at
elucidating the fundamental physics providing cells with their unique
material properties.  We make use of the fact that the mechanical
response of the cell can be ascribed to the cytoskeleton
\cite{Alberts2002}, a complex biopolymer network. Within the
constituents of the cytoskeleton the dominant response is contributed
by the actin cortex, a semiflexible polymer network connected by weak,
reversible crosslinks. It is the microstructure of this actin cortex
that we use as starting point for modeling. More precisely, we base
our discussion on a model for transiently crosslinked biopolymer
networks, called the inelastic glassy wormlike chain (inelastic \gwlc)
\cite{Wolff2010}. From the inelastic \gwlc, which is numerically
still quite demanding, we extract a reduced constitutive model that is
more tractable both analytically and numerically. We restrict
ourselves to reversible (``no-slip'') on-off kinetics, but find the
model nevertheless capable of describing a large part of the known
experimental data. For illustration, we evaluate the response to a
standard deformation protocol.

\section{The inelastic \gwlc}
The inelastic glassy wormlike chain (inelastic \gwlc)
\cite{Wolff2010} is a phenomenological mean-field type model that
describes a test polymer fluctuating against a background network. The
slowing-down of the conformational dynamics of the test polymer by the
surrounding network is phenomenologically represented by a stretching
of its mode spectrum. More precisely, the single mode relaxation time
$\tau_n$ of the $n$th eigenmode of a free polymer is modified
according to
\begin{equation}
  \tau_n \to 	\tilde{\tau}_n=\left\{ \begin{array}{c c} \tau_n &
\lambda_n < \Lambda \\ 
	\tau_n \exp\left[ \eps(\lambda_n/\Lambda -1)\right] & \lambda_n
\geq \Lambda \end{array}\right. .
\label{eq:gwlc}
\end{equation}
In the spirit of a mean-field model, $\eps$ and $\Lambda$ are below
interpreted as the characteristic free energy barrier of, and the average
contour distance between, weak transient bonds, respectively, and $\lambda_n$ is
the wavelength of the $n$th mode of the polymers's transverse contour
fluctuations $r_\perp(s,t)$. Inserting the modified relaxation spectrum,
Eq.~(\ref{eq:gwlc}), into the expression for the mechanical response of an
isolated wormlike chain in solution yields a
highly successful phenomenological parametrization of the equilibrium response
of sticky biopolymer networks \cite{Semmrich2007} and even of living cells
\cite{Kroy2009}. For example, the susceptibility of a test polymer of bending
rigidity $\kappa$ to a transverse point force is given by \cite{Kroy2007}
\begin{equation}
     \alpha(\omega)=\frac{\Lambda^3}{\kappa
\pi^4}  \int_{0}^{\infty} d\tilde n \frac1{(\tilde n^4+\tilde n^2
f/f_\Lambda)(1+i\omega
\tilde\tau_{\tilde n})},
  \label{eq:GWLC_mode_sum}
\end{equation}
where $f_\Lambda$ is the Euler buckling force.

We found the above interpretation of the mathematical expression in terms of
physical network parameters particularly useful for non-equilibrium processes,
in which the bond network is driven out of equilibrium so that the average bond
distance $\Lambda$ deviates from its force-free reference value $\Lambda_0$
according to
\begin{equation}
 \frac\Lambda{\Lambda_0}= \frac{\nu_0}\nu,
\end{equation}
where $\nu$ quantifies the fraction of broken bonds and $\nu_0$ is the
respective equilibrium value. For simplicity, we model $\nu$ by a first-order
kinetic equation with two possible states, bound and unbound, characterized by
on and off rates $k_+$ and $k_-$, respectively \cite{Wolff2010}
\begin{equation}
 \dot \nu(t,f) =-\left\{k_+(f) + k_-(f)\right\}\nu(t)+k_+(f).
  \label{eq:bond_kinetics}
\end{equation}
The rates are prescribed by Kramers` theory \cite{Kramers1940} and an
exponential force dependence according to Bell's model \cite{Bell1978}, {\it
i.e.\ }
\begin{equation}
    k_+(f)=k e^{U-\Delta x_u  f},
  \label{eq:kon}
\end{equation}
and
\begin{equation}
    k_-(f)=k e^{\Delta x_b  f},
  \label{eq:koff}
\end{equation}
with $k\sim\tau_0^{-1} e^{-\eps}$.
They depend on the height $\eps k_{\rm B}T$ of the the energy barrier
separating  bound and unbound state, the relative binding affinity $e^U$, and
the widths $\Delta x_b$ and $\Delta x_u$ of the bound and unbound state,
respectively. In the following, the thermal energy $k_{\rm B} T$ is set to one
for convenience.

\section{Simplified constitutive equations} 
The crucial feature of the inelastic \gwlc\ is that the mutual
interaction of the ``stiff'' polymer response and the ``soft'' bond
network generates a huge variety of phenomena, depending on the
particular stimulus. Here, we are interested in the long-time
quasi-plastic response, and not in short-time effects and the
particular shape of the relaxation spectrum, which we discussed
elsewhere \cite{Wolff2010}. We therefore refrain from reproducing all
the details of the full model and concentrate on those aspects that
are most important in this respect, namely a nearly exponential strain
stiffening and the softening due to the reversible breaking of weak
bonds. This leads to a reduced simplified formulation of the inelastic
\gwlc\ that allows for the {\em analytical derivation} of constitutive
equations for transiently crosslinked biopolymer networks.

To arrive at these equations, we cast the model equations into a form that
emphasizes common aspects of plasticity theory. As a
starting point, the system size $L$ is related to the initial size $L_0$ by a
scaling function $\alpha^{\rm nl}$ that characterizes the nonlinear compliance
in terms of the dimensionless bond fraction $\nu$,
\begin{equation}
 L= L_0 \alpha^{\rm nl}\left[f,\nu(f)\right].
\label{eq:current_length}
\end{equation}
The inelastic \gwlc\ model predicts a monotonic decrease of the material
stiffness with the bond fraction \cite{Wolff2010}, {\em i.e.\ }the more
bonds are broken, the more susceptible the material becomes to deformations. For
the sake of the argument, we approximate the precise functional form by a simple
reciprocal dependence of the nonlinear susceptibility on the bond fraction,
\begin{equation}
  \alpha^{\rm nl}\left[f,\nu(f)\right]\equiv \frac{\nu_0}{\nu} \delta(f,\nu_0).
\end{equation}
Here, we have introduced the (nonlinearly) elastic component $\delta$. It
only depends on the equilibrium bond fraction $\nu_0$ under zero external force,
while the dependence on the current bond fraction $\nu$ is isolated into the
prefactor. Defining the {\em inelastic length} $\mathcal L$ as the part of the
extension that is not due to elastic deformations, 
\begin{equation}
 \mathcal{L}\equiv L_0 \nu_0/\nu,
\label{eq:inelastic_length}
\end{equation}
the actual system length can now be written as 
\begin{equation}
  L = \mathcal{L} \cdot \delta.
\label{eq:L_decomp}
\end{equation}

Stressing the similarities to plasticity theory, we can associate a ``rest
force'' $\mathcal F$ with the inelastic length, defined by the condition
$\mathcal{\dot L}=0$. Given our knowledge of the bond kinetics, the rest force
is easily calculated. According to Eq.~(\ref{eq:inelastic_length}), 
$\mathcal{\dot L}=0 \Leftrightarrow \dot\nu=0$. The rest force $\mathcal F$ is
therefore determined by the following condition
\begin{equation}
 \mathcal{F}:\dot \nu(t)\vert_{\mathcal{F}} =0. 
\end{equation}
By Eqs.~(\ref{eq:bond_kinetics})-(\ref{eq:koff}), for a given
force $f$, $\dot\nu = 0 \Leftrightarrow \nu = \left(1+e^{(\Delta x_b+\Delta
x_u)f -U}\right)^{-1}$, yielding
\begin{equation}
 (\Delta x_b+\Delta x_u) \mathcal{F} = 
U+\ln\left(\frac{\mathcal L}{L_0 \nu_0}-1\right),
\label{eq:rest_force}
\end{equation}
where we solved for f and replaced $\nu$ using
relation~(\ref{eq:inelastic_length}).
In the context of bond kinetics, $\mathcal F$ is the force that would have to
be maintained to render the {\em current} bond fraction stationary. 
Taking a time derivate, the rest force is seen to inherit its dynamics from the
inelastic deformation rate $\dot {\mathcal L}$, a phenomenon commonly denoted as
kinematic hardening. However, in contrast to the common linear kinematic
hardening ($\dot {\mathcal F}\propto \dot{ \mathcal L}$), bond
breaking leads to the phenomenology of {\em logarithmic kinematic hardening},
$\dot{\mathcal F}\propto \dot{\mathcal L}/\mathcal L$. Note that $\mathcal F$ is
finite for ${\mathcal L} > L_0 \nu_0$ and that, by definition, $\mathcal F=0$
for $\mathcal L = L_0$.

The corresponding flow rule can be found by decomposing the total mechanical
force $f= \mathcal F+\Delta f$ into the rest force $\mathcal F$ and an
overstress $\Delta f$, and inserting it into the force-dependent
bond kinetics, Eqs.~(\ref{eq:bond_kinetics})-(\ref{eq:koff}). Eliminating
$\mathcal F$ using Eq.~(\ref{eq:rest_force}), and expressing the result in
terms of the inelastic length $\mathcal L$, the flow rule is given by
\begin{equation}
    e^{-\epsilon \cdot U} \cdot \tau_0 e^\eps \cdot
\frac{\dot{\mathcal{L}}}{\mathcal L} \cdot
\left(\frac{\mathcal L}{L_0 \nu_0}-1\right)^{-\epsilon} = \left(e^{\Delta x_b
\Delta f}-e^{-\Delta x_u \Delta f} \right),
\label{eq:flow_rule}
\end{equation}
where we introduced the abbreviation $\epsilon \equiv \Delta x_b/(\Delta
x_b+\Delta x_u)$. At any instant $t$, Eq.~(\ref{eq:flow_rule}) uniquely
determines $\mathcal L$, given the overstress history $\Delta f(t)$. Note that
the equations derived above do not depend on the particular choice of the
elastic response $\delta$.

To summarize this section, we found a set of constitutive equations from 
an approximate model based on the molecular structure of the cytoskeleton,
describing the inelastic mechanical response of a transiently
crosslinked biopolymer network. We can now proceed to evaluate the equations
and to numerically examine the responses to a simple deformation protocol.

\section{Quasi-plastic response}
\begin{figure}
\includegraphics{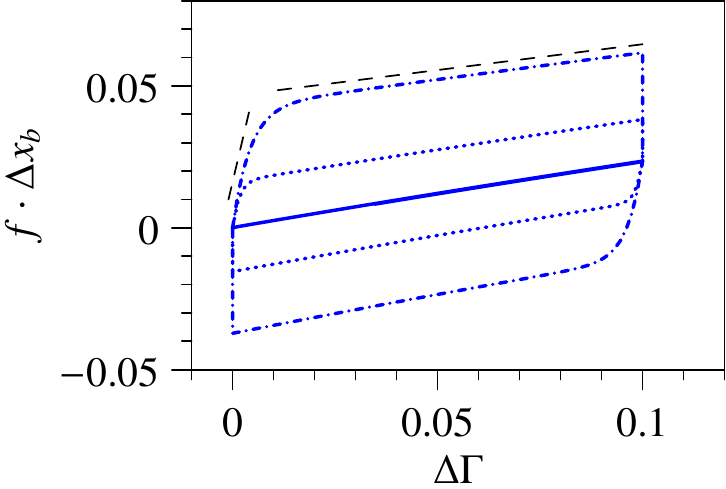}
\includegraphics{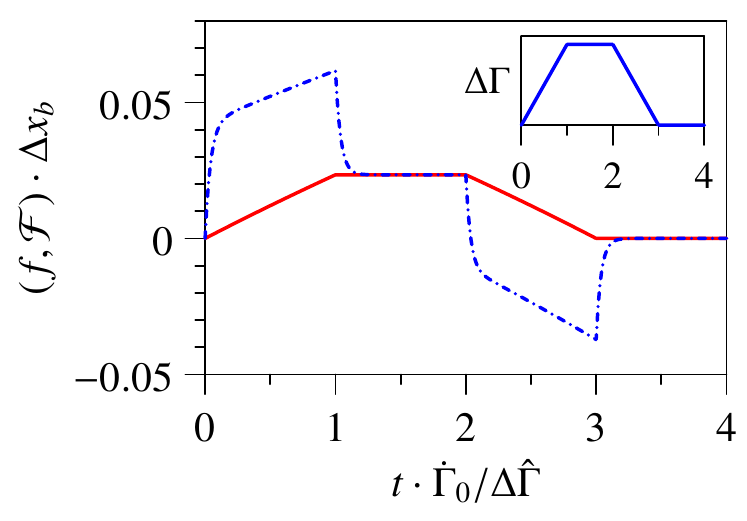}
\caption{Response of the simplified schematic model for biopolymer
  networks, Eqs.~(\ref{eq:L_decomp})-(\ref{eq:flow_rule}),
  derived from the inelastic \gwlc, to a piece-wise linear deformation
  protocol (loading-waiting-unloading, see inset of lower panel). The
  relative deformation amplitude is $\Delta \hat \Gamma=10\%$, and
  ramp and plateau durations are $\Delta \hat
  \Gamma/\dot{\Gamma}_0$. \emph{Upper panel:} force-displacement
  curves [deformation rates $\dot \Gamma_0 \tau_0 e^\eps=$ 0.004
  (solid lines), 0.64 (dotted lines), and 1.6 (dot-dashed lines)];
  dashed lines indicating elastic and apparent kinematic hardening
  regimes, respectively, reminiscent of what is observed for living 
   cells \cite{Fernandez2008}. \emph{Lower panel:} time-dependent force for
  the highest deformation rate (dot-dashed lines) compared to the
  time-dependent rest force $\mathcal F$ (solid lines); forces are
  normalized by the natural force scale $\Delta x_{\rm b}^{-1}$ (in
  units of $k_{\rm B}T$) of the bound state. }
  \label{fig_th_const_single}
\end{figure}

We consider the response of the model to a deformation ramp $\dot
\Gamma \equiv \dot L/L_0= \dot \Gamma_0\equiv{\rm const.}$, followed
by a plateau $\dot \Gamma = 0$ and an inverse ramp, $\dot \Gamma =
-\dot \Gamma_0$ (Fig.~\ref{fig_th_const_single}, lower panel, inset). The
elastic contribution of the inelastic \gwlc\ is well approximated by
exponential elasticity, $f(\delta)\propto (\delta-1) e^{\gamma \cdot
  (\delta-1)^2}$, which has been observed for a multitude of
biomaterials \cite{Fung1993}.  For simplicity, we also neglect the
viscous component of the viscoelastic polymer response, as it is not
essential for the following argumentation. The qualitative effects
presented in the following do not depend on these technically
motivated simplifications.

The force-displacement curves exhibit rate-dependent hysteresis
(Fig. \ref{fig_th_const_single}, upper panel), which in our model is
a signature of dissipation caused by inelastic reversible bond
breaking. In the limit of an infinitely slow deformation, the bonds
are always in equilibrium and the hysteresis vanishes (solid lines in
the upper panel of Fig.~\ref{fig_th_const_single}), since we
dismissed the {\em viscoelastic} hysteresis present in the \gwlc\
model and real biopolymer materials, here. For low to moderate rates
(the time scale is set by the intrinsic time scale $\tau_0 e^\eps$ of
the bonds), all force-displacement curves share characteristic
features. Most prominently, the response to a linear ramp is
characterized by two approximately linear regimes, emphasized by the
dashed lines in the upper panel of
Fig.~\ref{fig_th_const_single}. The first linear regime can be
interpreted as an elastic response. The second regime is explained by
a quasi-plastic deformation with slowly increasing rest force
$\mathcal F$, a signature of the apparent kinematic hardening
identified above [see Eq.~(\ref{eq:rest_force})], where we found
that the rest force depends {\em logarithmically} on the inelastic
deformation, suggesting the notion of ``logarithmic kinematic
hardening''. Corresponding deviations from the linear
force-displacement curve become discernible for deformations with
amplitudes much larger than $10\%$ (not shown). The phenomenology 
is reminiscent of experimental results obtained for living cells \cite{Fernandez2008}.

The interpretation of an elastic and an inelastic regime is substantiated by
comparing the time-dependent rest force $\mathcal F$ to the total mechanical
force $f$ (Fig.~\ref{fig_th_const_single}, lower panel). While during a ramp, the total
mechanical force initially strongly diverges from the rest force, it
quickly settles on a course parallel to the rest force, consistent with a
constant elastic contribution. The elastic contribution is relaxed upon halting
the deformation. The force characterizing the transition between elastic and
inelastic regime can be interpreted as an effective yield threshold. For
infinitely slow deformations, the total mechanical force equals the rest force,
and no predominantly elastic regime is present.  In other words, also the
effective yield threshold depends on the deformation rate. This observation sets
our biopolymer network apart from usual models for hard solids. The reason for
this behavior is that, by construction, the bonds will always yield if the
stimulus is slow compared to the bond-opening time scale. Only for sufficiently
fast deformation, an initial predominantly elastic response can be obtained.

\section{Conclusions} 
Starting from a minimal model of a transiently crosslinked biopolymer
network, we derived quasi-plastic constitutive equations exhibiting
logarithmic kinematic hardening. In contrast to ``truly'' plastic
materials, the inelastic deformations are rooted in the {\em
  inelastic, reversible softening} due to the transient breaking of
weak bonds, as presumed by the cell rheological model of the inelastic
\gwlc.

From the perspective of biomechanics, the present work may lead to an
intuitive understanding of the mechanical properties of biopolymer
materials and cells.  It also provides a simple but accurate model to
simulate the large-scale behavior of the materials, {\it e.g.\ }using
finite element methods. From the perspective of materials science, our
work sheds light on a new class of materials, which can bear a kind of
strain that is both recoverable {\em and} dissipative, as opposed to
the ``usual'' reversible elastic and the irreversible plastic
strain. This recoverable inelastic strain bears resemblance to the
quasi-plastic-elastic (QPE) model recently proposed in relation with
DP steel alloys \cite{Sun2011}. In contrast to the QPE strain,
however, our recoverable inelastic strain is rate-dependent, due to
the underlying slow bond dynamics. It is an intriguing question
whether the QPE strain might emerge from the bond-breaking approach in
some special limiting case. As an outlook, we would like to mention
that it would be straightforward to extend our model to account for
true plastic strains \cite{Fernandez2008} (by associating some slip
with each bond breaking event) as well as for the
physiologically important internally generated active stresses
\cite{Wang2002, Stamenovic2004}.

\begin{acknowledgments}
We are grateful to Pablo Fern\'andez for helpful discussions and acknowledge financial support from the German excellence initiative via the Leipzig School of Natural Sciences - Building with Molecules and Nano-objects (BuildMoNa). 
\end{acknowledgments}


\bibliography{/home/wolff/Uni/library}

\end{document}